\documentclass{aa}

\usepackage{times}
\usepackage{graphicx}

\newcommand{\be}{\begin{equation}}
\newcommand{\ee}{\end{equation}}
\newcommand{\bdm}{\begin{displaymath}}
\newcommand{\edm}{\end{displaymath}}
\newcommand{\bea}{\begin{eqnarray}}
\newcommand{\eea}{\end{eqnarray}}

\newlength{\figheight}

\newcommand{\VV}[1]{\mbox{\boldmath{
                    ${\rm #1}$}}}

\newcommand{\dif}[2][]{\ensuremath{\/
                       \frac{\partial\/#1\/}%
                            {\partial\/#2\/}\/}}

\newcommand{\Dif}[2][]{\ensuremath{\/
                       \frac{\mathrm{d}\/#1\/}%
                            {\mathrm{d}\/#2\/}\/}}       

\newcommand{\nab}[0]{\VV{\nabla}\/}           

\newcommand{\Dim}[1]{\ensuremath{
                     \/\mathrm{#1}\/}}

\newcommand{\invh}[1]
   {\newlength{\hei}%
    \newlength{\dep}%
    \settoheight{\hei}{#1}%
    \settodepth{\dep}{#1}%
    \addtolength{\hei}{\dep}%
    \rule[\dep]{0pt}{\hei}%
   }
\newcommand{\invw}[1]
   {\newlength{\wid}%
    \settowidth{\wid}{#1}%
    \rule{\wid}{0pt}%
   }

\begin{document}

\thesaurus{11.14.1, 11.13.2, 02.13.2, 02.01.1}

\title{Magnetic Reconnection and Particle Acceleration in \\
       Active Galactic Nuclei}
\titlerunning{Magnetic Reconnection and Particle Acceleration in 
              Active Galactic Nuclei}

\author{R.~Schopper \and H.~Lesch \and G.T.~Birk}

\offprints{R.~Schopper}

\institute{%
  Institut f{\"u}r Astronomie und Astrophysik der
  Universit{\"at} M{\"u}nchen, Scheinerstra{\ss}e 1, D-81679
  M{\"u}nchen, Germany
}

\date{Received date ; accepted date}

\maketitle

\markboth{Particle simulation studies of acceleration processes
in active galactic nuclei}
{Schopper et al.}

\begin{abstract}
Magnetic field-aligned electric fields are characteristic features of
magnetic reconnection processes operating in externally agitated
magnetized plasmas. An especially interesting environment for such a
process are the coronae of accretion disks in active galactic nuclei
(AGN). There, Keplerian shear flows perturb the quite strong disk
magnetic field leading to intense current sheets.  It was previously
shown that given field strengths of 200 G in a shear flow,
reconnection driven magnetic field aligned electric fields can
accelerate electrons up to Lorentz factors of about 2000 in those
objects thus providing us with a possible solution of the injection
(pre-acceleration) problem.  However, whereas in the framework of
magnetohydrodynamics the formation of the field-aligned electric
fields can be described consistently, the question has to be addressed
whether the charged particles can really be accelerated up to the
maximum energy supplied by the field-aligned electric potentials,
since the accelerated particles undergo energy losses either by
synchrotron or inverse Compton mechanisms. We pre\-sent relativistic
particle simulations starting from electric and magnetic fields
obtained from magnetohydrodynamic simulations of magnetic reconnection
in an idealized AGN configuration including nonthermal radiative
losses.  The numerical results prove that the relativistic electrons
can be effectively accelerated even in the presence of an intense
radiation bath.  Energies from~$50 \Dim{MeV}$ up to~$40 \Dim{GeV}$ can
be reached easily, depending on the energy density of the photon
bath. The strong acceleration of the electrons mainly along the
magnetic field lines leads to a very anisotropic velocity distribution
in phase space. Not even an extremely high photon energy density is
able to completely smooth the anisotropic pitch angle distribution
which is characteristic for quasi monoenergetic particle beams.
\end{abstract}


\hfill

\section{Introduction} \label{sec:intro}

Active galactic nuclei (AGN) can be regarded as accreting supermassive
black holes surrounded by accretion disks (Camenzind 1990; Miyoshi et
al. 1995; Burke and Graham-Smith 1997 and references therein).
Relativistic electrons in AGN reveal themselves by hard X-ray
(probably due to pair production (cf. Svensson 1987; Done and Fabian
1989)) and $\gamma$-ray emissions as well as radio observations of
superluminous motions (e.g. Abraham et al. 1994).  The
$\gamma$-radiation observed from quasars and bla\-zars may originate
in a distance $R$ of $10^{2-3}$ gravitational radii from the central
engine (Dermer and Schlickeiser 1993).  At that distance no
significant pair production happens but relativistic leptons scatter
via the inverse Compton process the IR-UV radiation of the accretion
disk within a relativistically moving jet.  It is well known that
``standard'' mechanisms for the acceleration of high energy leptons,
as diffusive shock wave acceleration and resonant acceleration by
magnetohydrodynamical (MHD) turbulence can only work efficiently for
Lorentz factors larger than $\gamma_{\rm crit}\simeq m_{\rm p}/{m_{\rm
e}}$ (where $m_{\rm p}$ and $m_{\rm e}$ denote the proton and electron
masses). Consequently, charged particles accelerated via shocks or
MHD~turbulence have to be pre-accelerated which confronts us with the
{\bf injection problem} (Blandford 1994; Melrose 1994) in the AGN
context.

In a differentially rotating magnetized accretion disk gas, the
evolution of a magnetized corona is quite hard to suppress (Galeev et
al. 1979; Stella \& Rosner 1984).  Driven by the buoyancy force
magnetic flux tubes ascend into the disk corona, thereby their
footpoints are sheared by the differential rotation of the
disk. Either by internal shear or by encountering already pre\-sent
magnetic flux, magnetic reconnection and accompanied rapid dissipation
of magnetic energy happens in the coronal plasma. Such a behavior can
be studied with great detail for example in the solar corona (Parker
1994 and references therein). In a recent contribution we investigated
the possible role of magnetic field-aligned electric fields
($E_\parallel $) in the context of magnetic reconnection operating in
AGN coronae for the pre-ac\-cel\-er\-ation of leptons (Lesch and Birk
1997; hereafter LB). It could be shown that field-aligned electric
potential structures in relatively thin current sheets form. For
reasonable physical parameters such electric potentials are strong
enough to accelerate electrons up to $\gamma \approx 2000$, in
principle. However, in the framework of MHD the actual energies of the
accelerated particles cannot be calculated.  It is the aim of the
present contribution to corroborate the model introduced in LB with
the help of relativistic particle simulations by taking macroscopic
electric and magnetic field configurations obtained by the MHD
simulations as an input.

In the next section we resume the MHD model in a nutshell and present
the details of the resulting three-dimensional electric and magnetic
fields.  In Sec.~\ref{sec:simulation} we discuss our approach to the
numerical study of high-energy particles and show the numerical
results dwelling on particle spectra and energies. Eventually, we
discuss our findings in Sec.~\ref{sec:Disc}.

\section{The MHD model for magnetic reconnection
in the corona of an accretion disk} \label{sec:MHD}

The MHD framework for the particle simulations is modeled by means of
a 3D resistive MHD code that integrates the one-fluid balance
equations (e.g. Krall and Trivelpiece 1973):
\be
{\partial \rho \over \partial t} + \nabla \cdot (\rho {\bf v}) = 0
\ee
\be
{\partial \over \partial t } (\rho {\bf v})
+ \nabla \cdot (\rho {\bf v} {\bf v}) = -\nabla p
        + \frac{1}{c} \, {\bf j} \times {\bf B}
\ee
\be
{\partial p \over \partial t} + \nabla \cdot (p {\bf v}) =
{2\over3}(-p  \nabla \cdot {\bf v}+\eta j^2)
\ee
\be
{\partial {\bf B} \over \partial t} = \nabla \times ({\bf v}\times {\bf B})
- \frac{c^2}{4 \pi} \nabla \times ( \eta {\bf j} )
\ee
where $\rho$, ${\bf v}$, $p$, ${\bf B}$, ${\bf j}$ and $\eta$ denote
the plasma mass density, velocity and pressure, the magnetic field,
the electric current density and the electrical resistivity.

The scenario we have in mind can be briefly described as follows (for
details see LB): Different convective plasma motions at different
regions of an AGN corona or differential shear motion of the disk
result in a sheared magnetic field and consequently, in the formation
of field-aligned electric current sheets.  When this convective shear
motion is strong enough to result in a supercritical current density
(drift velocity larger than the local sound velocity $c_{\rm
s}=\sqrt{k_{\rm B} T_{\rm e}/m_{\rm p}}$ current-driven
microinstabilities are excited which lead to anomalous electrical
resistivity (e.g. Huba 1985; Benz 1993).  The formation of localized
regions of anomalous dissipation or, more general, the local violation
of ideal Ohm's law, can be regarded as the necessary onset condition
for magnetic reconnection (Schindler et al. 1988, 1991).  During the
dynamical evolution of the reconnection process very localized
acceleration regions (${\bf E}_{\vert \vert} \ne 0$) form.
\begin{figure}[htbp]
  \centering
  \includegraphics[angle=0,width=\hsize,totalheight=\figheight,keepaspectratio]{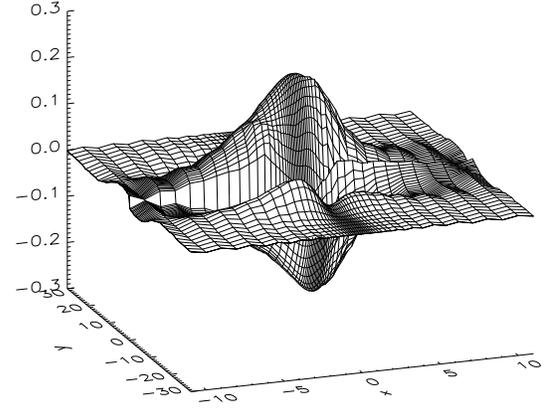}
  \includegraphics[angle=0,width=\hsize,totalheight=\figheight,keepaspectratio]{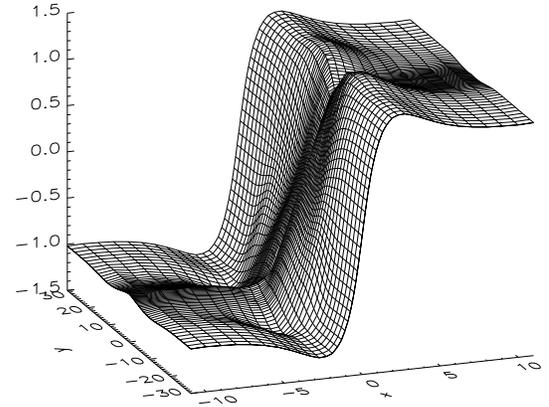}
  \includegraphics[angle=0,width=\hsize,totalheight=\figheight,keepaspectratio]{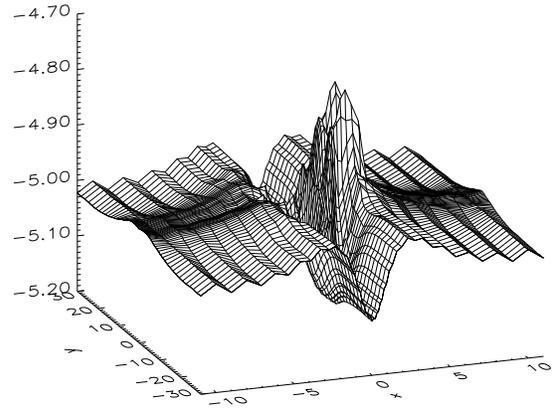}
  \caption[]{The $x$-,$y$- and $z$-component (from top to bottom) of
  the reconnection magnetic field at the height of the central
  reconnection region after $t=120\tau_{\rm A}$.}
  \label{fig:B-Comp}
\end{figure}
\begin{figure}[htbp]
  \centering
  \includegraphics[angle=0,width=\hsize,totalheight=\figheight,keepaspectratio]{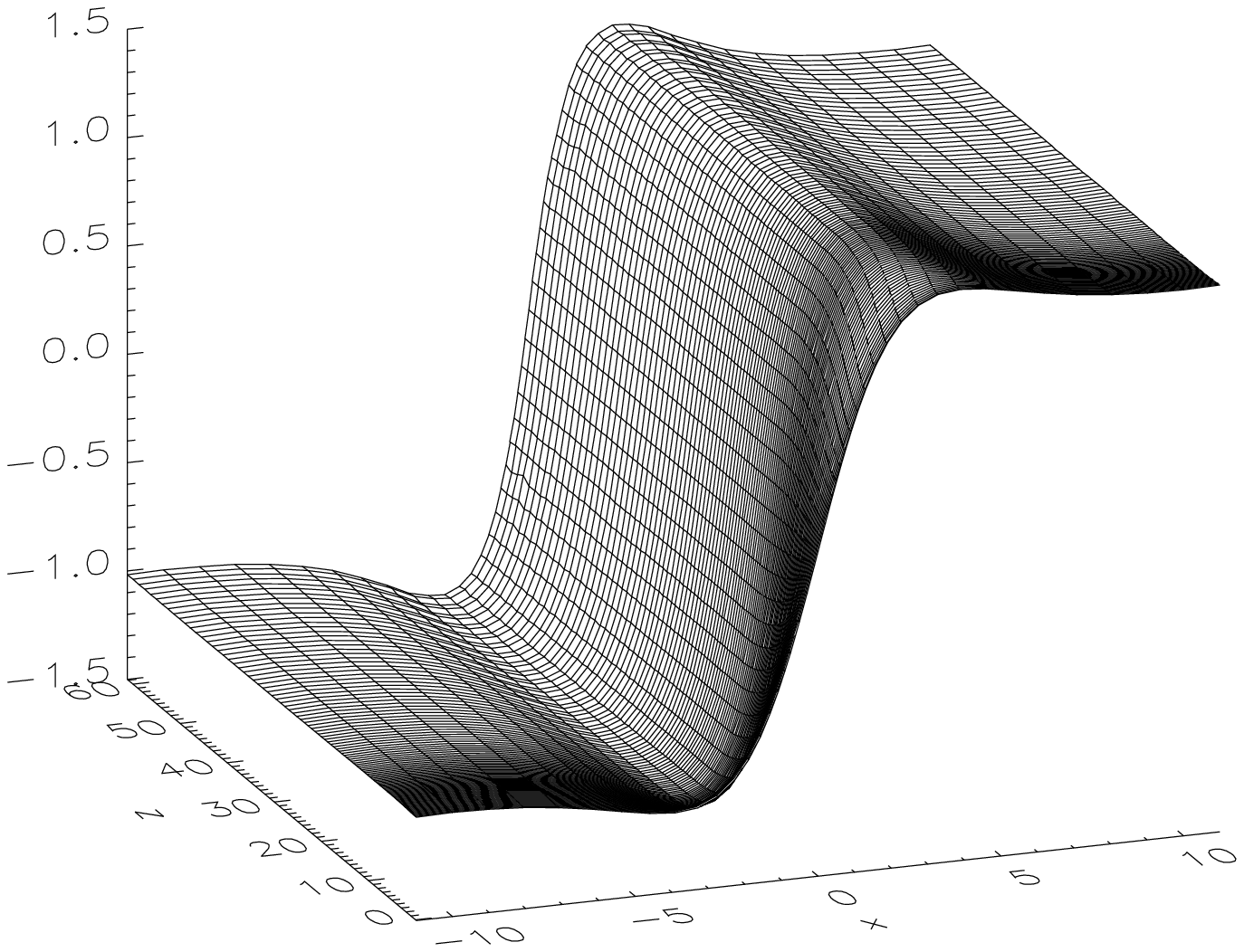}
  \includegraphics[angle=0,width=\hsize,totalheight=\figheight,keepaspectratio]{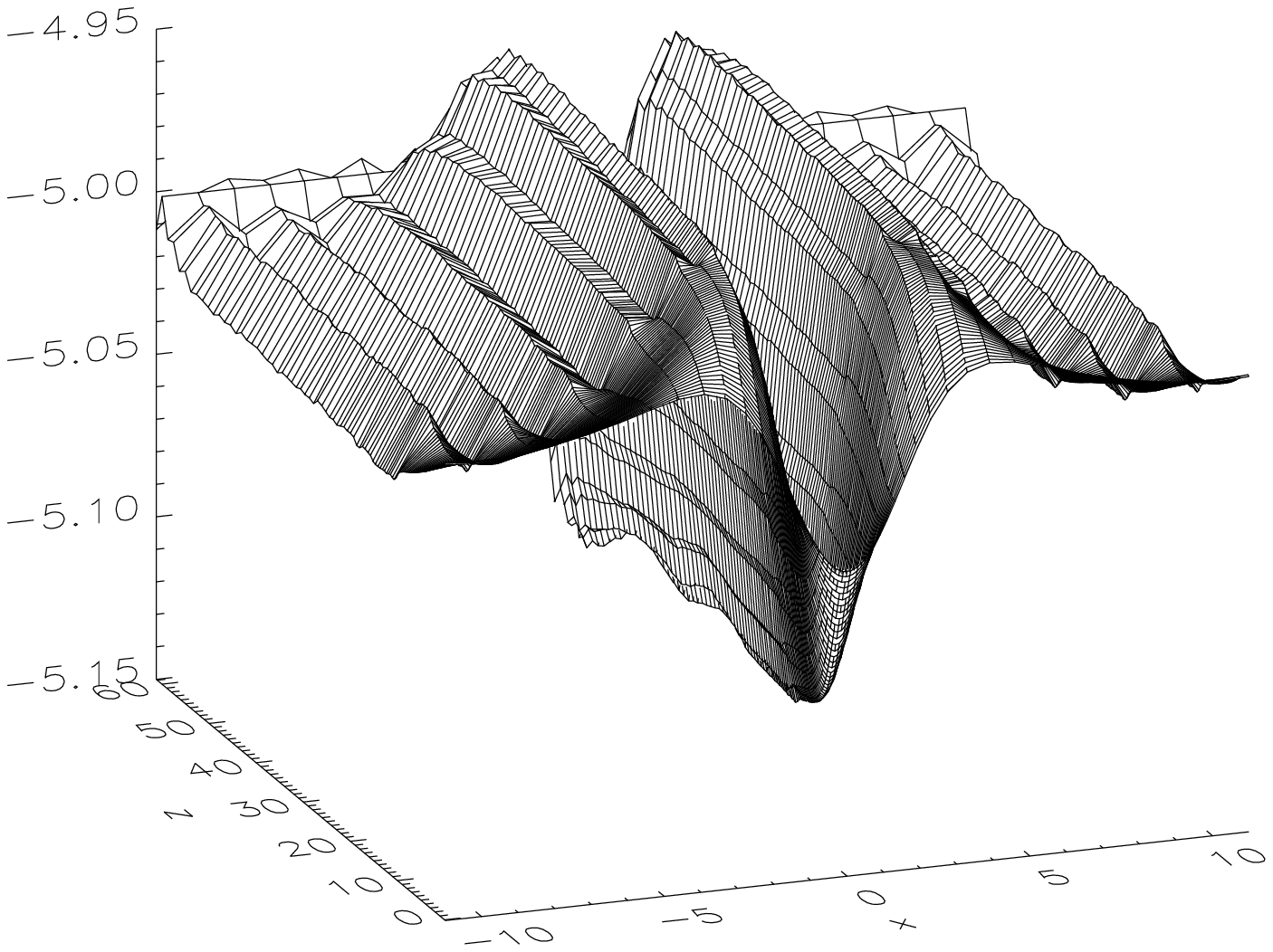}
  \caption[]{The $y$- and $z$-component of the reconnection magnetic
  field at the $y=0$-half plane (i.e. the middle on the of the central
  reconnection region) after $t=120\tau_{\rm A}$.}
  \label{fig:B-Field}
\end{figure}
\begin{figure}[htbp]
  \centering
  \includegraphics[angle=0,width=\hsize,totalheight=\figheight,keepaspectratio]{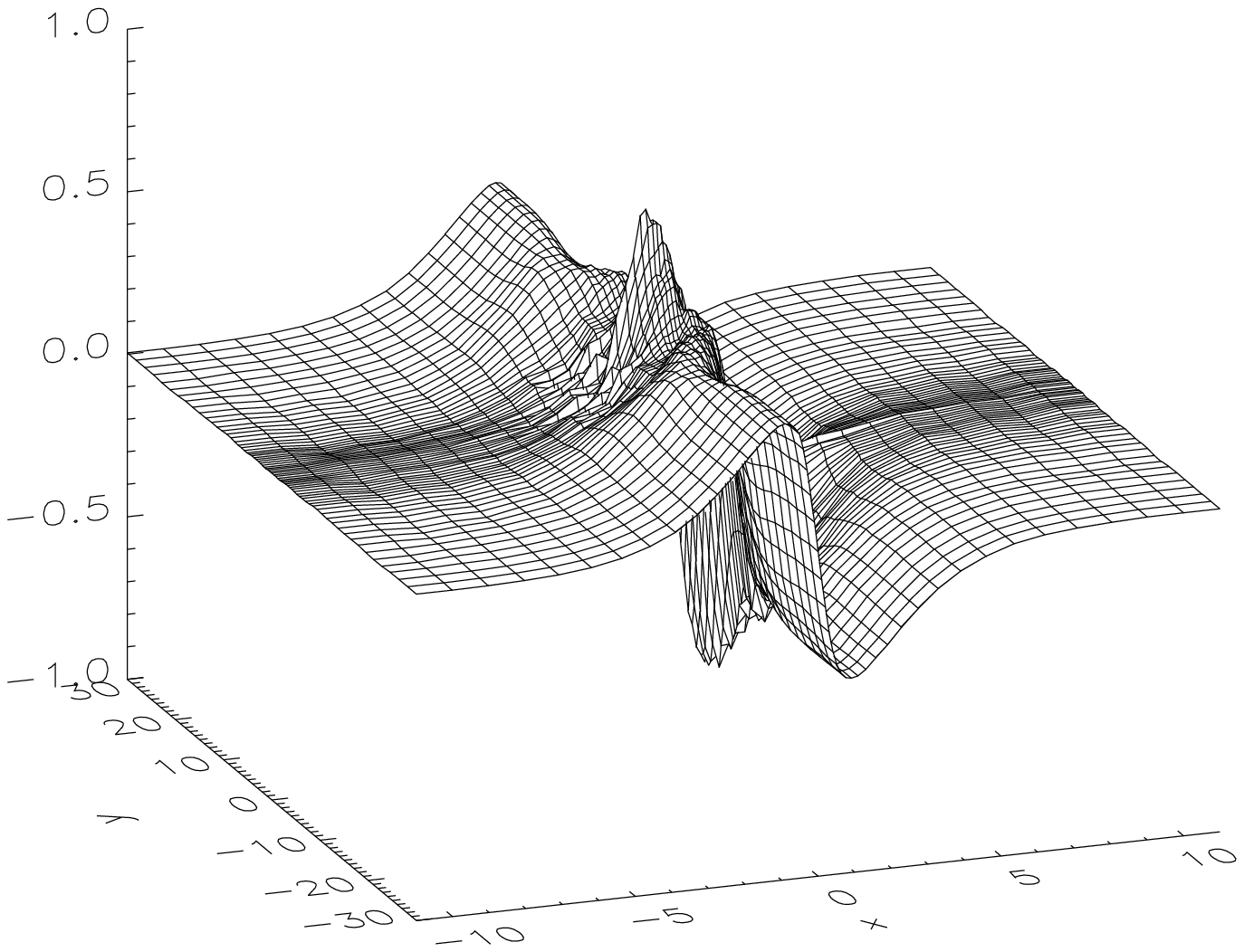}
  \includegraphics[angle=0,width=\hsize,totalheight=\figheight,keepaspectratio]{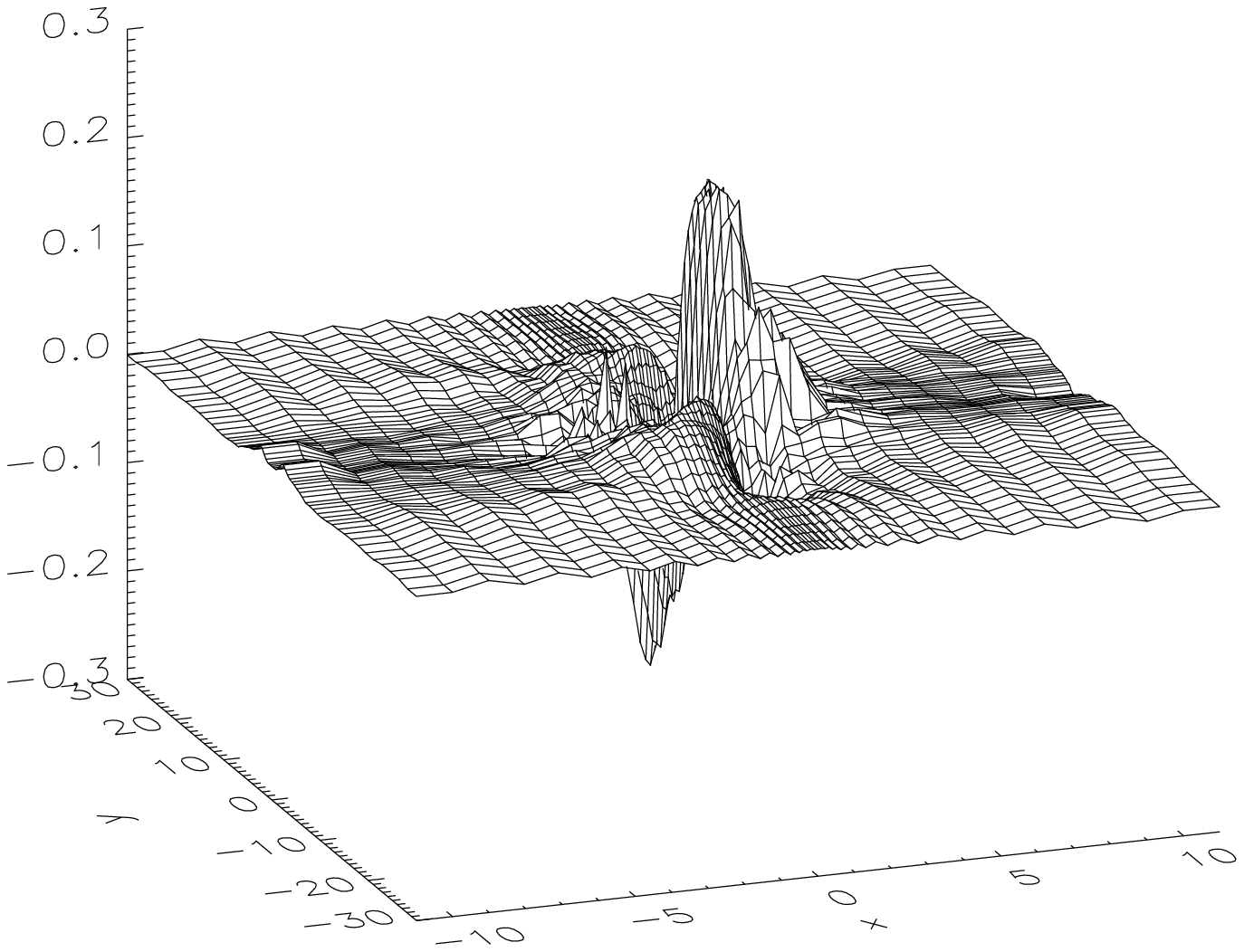}
  \includegraphics[angle=0,width=\hsize,totalheight=\figheight,keepaspectratio]{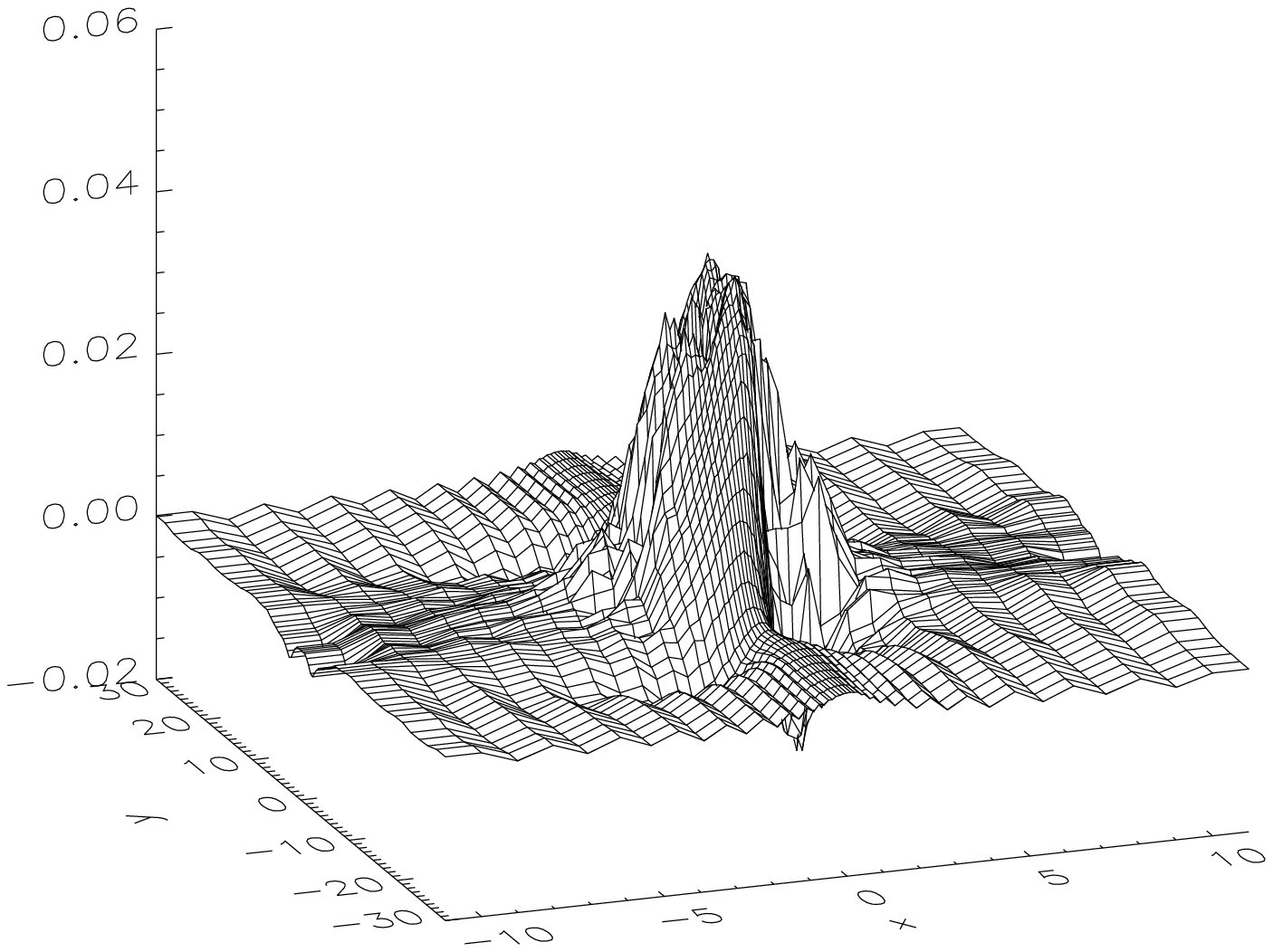}
  \caption[]{The $x$-,$y$- and $z$-component (from top to bottom)
  of the reconnection electric field after $t=120\tau_{\rm A}$.}
  \label{fig:E-Comp}
\end{figure}
\begin{figure}[htbp]
  \centering
  \includegraphics[angle=0,width=\hsize,totalheight=\figheight,keepaspectratio]{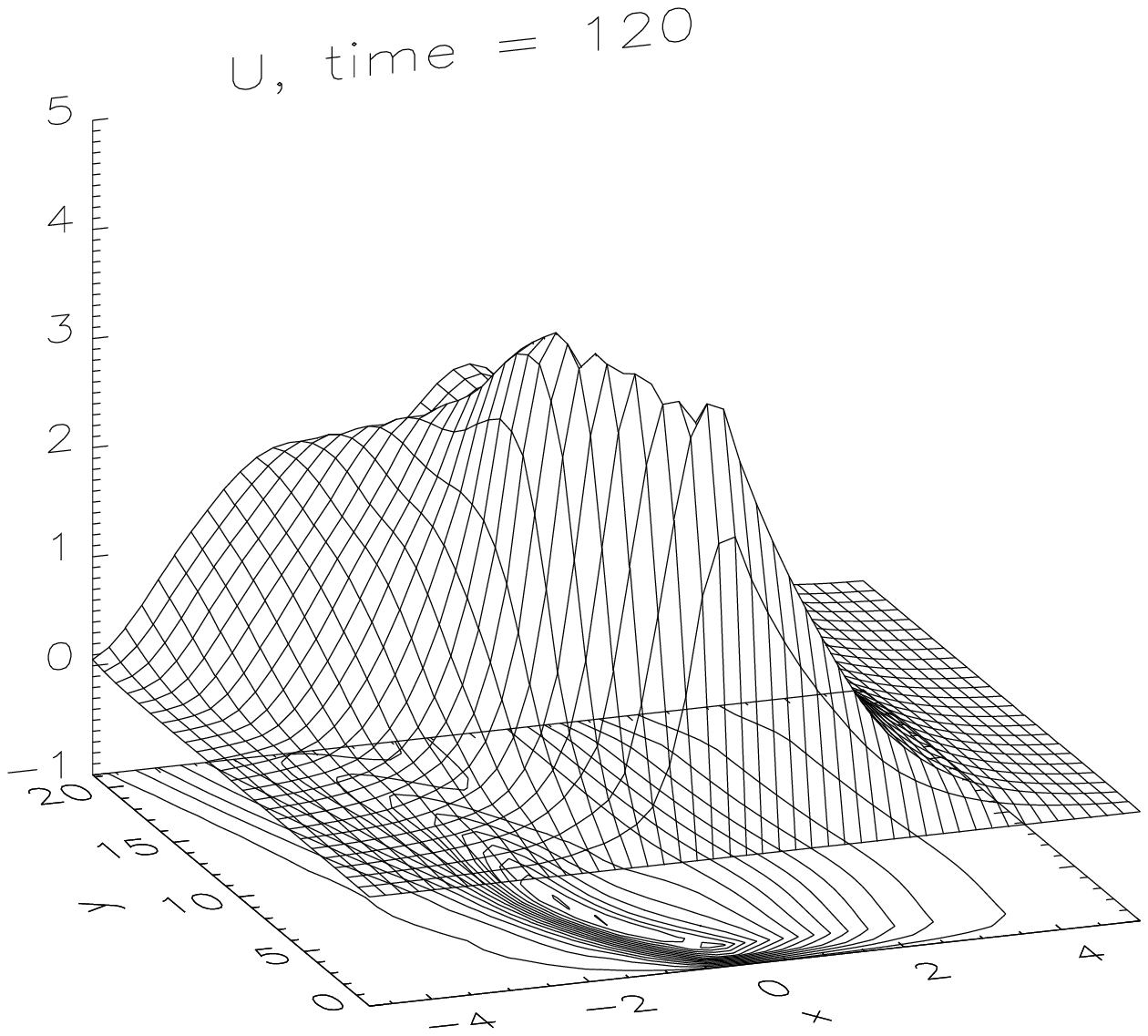}
  \caption[]{The generalized electric potential $U$ after
  $t=120\tau_{\rm A}$.}
  \label{fig:Potential}
\end{figure}

Since we want to concentrate on a single sheared coronal loop which is
represented by a current-carrying magnetic flux tube we start from an
appropriate idealized initial configuration for the MHD simulation run
which is characterized by a homogeneous plasma and a force-free
magnetic field (cf. Birk and Otto 1996; LB):
\be
{\bf B} = B_{y0}\tanh(x)\;{\bf e}_y\ - \sqrt{{B_{z0}}^2 +
{{B_{y0}}^2\over\cosh^2(x)}}\;{\bf e}_z
\ee
where $B_{z0}=5$ and $B_{y0}=1$ denote the normalized constant main
component and the shear (toroidal) component of the magnetic field,
respectively. As discussed in LB we choose for the half-width of the
current sheet $w \approx 10^7 {\rm cm}$, a shear magnetic field of
$B_{y0} = 200 {\rm G}$, a particle density of $n= 3 \cdot 10^5
{\rm cm}^{-3}$ and $10^4{\rm statamp\,cm}^{-2}$ for the critical
current density of $j_{\rm crit}$, which corresponds to an electron
temperature of $6\cdot 10^7\, K$ (cf. Ulrich 1991; Nandra and Pounds
1994). The magnetic Rey\-nolds num\-ber is chosen as $S = 4 \pi v_{\rm
A} w / \eta c^2 = 10^3$.

The initial equilibrium configuration is perturbed by some sheared
flow:
\be
v_y(x,z,t)=v_{y0}{\tanh(2x)\over\cosh^2\left({x\over 3}\right)}
\mathrm{e}^{-{z \over 6}}
\ee
with an amplitude chosen as $v_{y0} = 0.5 \%$ of the shear Alfv\'en
velocity $v_{\rm A} = B_{y0}/\sqrt{4 \pi \rho}$. This perturbation is
transported down along the main component of the magnetic field via
shear Alfv\'en waves during the dynamical evolution. An anomalous
resistivity will be switched on when the current density exceeds the
critical value $j_{\rm crit}$. The resistivity is localized in height
along the poloidal component of the magnetic field in order to model
an acceleration region of the length of $40 w$ as well as in the
$y$-direction (for technical details we again refer to LB).  During
the dynamics reconnection electric and magnetic fields evolve which
can be used as an input configuration for the particle simulation of
the acceleration of high-energy electrons in the reconnection
region. A snapshot after $t=120$ dynamic times (i.e. $\tau_{\rm A}=
w/v_{\rm A}$) of the Cartesian components of the magnetic and electric
fields are shown in Figures \ref{fig:B-Comp}--\ref{fig:E-Comp}. The
resulting $x$-component of the magnetic field shows the typical
bipolar reconnection characteristic (Fig.~\ref{fig:B-Comp}, top
panel). The middle panel of Fig.~\ref{fig:B-Comp} shows the
$y$-component of the magnetic field associated with the current sheet
and the lower panel shows the $z$-(poloidal) component.
Fig.~\ref{fig:B-Field} shows the $y$- and $z$-component of the
magnetic field at the $y=0$-half plane. The $y$-component is enhanced
locally due to the sheared flow applied at the upper boundary and
transported along the poloidal component of the magnetic field via
shear Alfv\'en waves.  The three-dimensional electric vector field at
the height of the central reconnection region is illustrated in
Fig.~\ref{fig:E-Comp}. The convective $x$-component (top panel) is the
result of the applied sheared flow and the divergent reconnection flow
whereas the convective $y$-component (middle panel) is due to the
convergent ($v_x(x)$) reconnection flow.  The dominant accelerating
component is $E_z$ (lower panel) which is caused by the finite
anomalous electrical resistivity $E_z \sim \eta j_z$ and accordingly
localized in the $x$- and $y$-direction.

The generalized field-aligned electric potential in normalized units
$U=-\int E_{\vert \vert} ds$ ($ds$ is a magnetic field line element)
associated with the electric and magnetic fields is illustrated in
Fig.~\ref{fig:Potential}.  It is this elongated relatively thin
potential structure by which the electrons are accelerated as will be
discussed in the next section.

\section{Particle simulations} \label{sec:simulation}

For the sake of simplicity, we perform our simulations as relativistic
test particle simulations. This means, that any interaction between
the simulated particles are neglected, as well as the backreaction of
the considered particles on the electromagnetic fields. In our
calculations we use the stationary fields, which are explained in
Sec.~\ref{sec:MHD}, simply as the "stage" for the accelerated charged
particles.

In addition to the acceleration force, we consider the radiative
losses in our simulations, especially the synchrotron~(SY) and the
inverse Compton~(IC)-losses. In particular, we are interested in the
question:

Is magnetic reconnection able to accelerate particles to relativistic
energies in the presence of synchrotron and inverse Compton losses?

\subsection{Momentum Equation and Numerical Procedure} \label{sub:numeric}

From the mathematical point of view the problem of test particle
simulation is an integration of the equations of motion for a single
charged particle. Since we are dealing with test particles, we have to
integrate:
\begin{eqnarray}
   \Dif[\VV{r}]{t} & = & \VV{v} \\                   \label{equ:motion1}
   \Dif[\VV{p}]{t} & = & \VV{F}(\VV{r},\VV{v},t) \;. \label{equ:motion2}
\end{eqnarray}
This simplifies our problem enormously. As usual, the
vectors~$\VV{r}$, $\VV{v}$ and $\VV{p}$ represent position, velocity
and momentum of the particle and $\VV{F}$ is the force acting on the
particle.  We use the relativistic relations
\begin{equation}
   \rule{0mm}{6mm}
   \VV{p} = \frac{m_0 \VV{v}}{\sqrt{1-\left( \displaystyle \frac{v}{c}
            \right)^2}}
   \quad\!\! \Longleftrightarrow \quad
   \VV{v} = \frac{\VV{p}}{m_0 \sqrt{ 1 + \left( \displaystyle
   \frac{p}{m_0 c} \right)^2 }} \label{equ:relativ}
\end{equation}
between velocity and momentum, with the rest mass of the
particle~$m_0$ and the velocity of light~$c$.

The resulting force acting on the particle is, in the simplest case,
given by the Lorentz force~$\VV{F}_{\rm L} = q \left[ \VV{E} + \VV{v}/c
\times \VV{B} \right]$, where $q$ denotes the electric charge of the
particle. This is used in calculations in which radiative loss
processes are neglected. If we take these losses into account, we
require in addition a radiative loss force, which describes the
corresponding interaction between radiation field (photons) and
particles. Such a force
\begin{eqnarray}
  \VV{F} & = & \frac{2 q^3 \gamma}{3 m c^3} \left\{ \!\!
               \rule{0mm}{5.6mm} \left( \! \dif{t} + \left( \VV{v}
               \! \cdot \! \nab \right) \! \right) \! \VV{E} +
               \frac{1}{c}\,\VV{v} \times \left( \! \dif{t} + \left(
               \VV{v} \! \cdot \! \nab \right) \! \right) \! \VV{B}
               \rule{0mm}{5.6mm} \right\} \nonumber \\
         & + & \frac{2 q^4}{3 m^2 c^4} \left\{ \rule{0mm}{5.6mm}
               \VV{E} \times \VV{B} + \frac{1}{c} \, \VV{B} \times
               \left( \VV{B} \times \VV{v} \right) + \frac{1}{c} \,
               \VV{E} \left( \VV{v} \cdot \VV{E} \right)
               \rule{0mm}{5.6mm} \right\} \nonumber \\
         & - & \frac{2 q^4 \gamma^2}{3 m^2 c^5} \, \VV{v} \left\{
               \rule{0mm}{5.6mm} \left( \VV{E} + \frac{1}{c} \, \VV{v}
               \times \VV{B} \right)^2 - \frac{1}{c^2} \left( \VV{E}
               \cdot \VV{v} \right)^2 \rule{0mm}{5.6mm} \right\}
\end{eqnarray}
can be found in Landau and Lifshitz (1951). It consists of
\emph{every} radiative loss process due to field--particle
interaction, including SY--~and IC--losses. Losses due to
particle--plasma--wave interaction are not considered up to now.

The force given in Landau and Lifshitz (1951) is used here in a
modified form. All terms with temporal or spatial derivatives are
neglected, because the external fields are varying on great scales and
it is very expensive to calculate those terms for the microscopic
photon bath. So we consider the IC--losses by including an additional
term, which describes the space- and time-averaged influence of the
IC--losses of an isotropic photon bath with the energy density~$U_{\rm
Rad}$. After some calculations this leads to the modified expression
\begin{eqnarray}
   \VV{F}_{\rm Rad} & \approx & \frac{2}{3} \frac{q^4}{m_0^2 c^4}
   \left\{ \VV{F}_{\rm L} \times \VV{B} + \VV{E} \cdot \left( \VV{E}
   \cdot \VV{\beta} \right) \rule{0mm}{5.6mm} \right. \nonumber \\
   & - & \left. \rule{0mm}{5.6mm} \gamma^2 \VV{\beta} \left[ F_{\rm L}^2
   + \frac{16}{3} \pi \beta^2 U_{\rm Rad} - \left( \VV{\beta} \cdot
   \VV{E} \right)^2 \right] \right\}. \label{equ:loss}
\end{eqnarray}
In this equation $\VV{\beta} = \VV{v}/c$ is the
velocity normalized to the speed of light.

The choice of numerical routines needs some comments. The routine,
which is executed most, is the evaluation of the force~$\VV{F} =
\VV{F}_{\rm L} + \VV{F}_{\rm Rad}$ at every timestep. This evaluation
consists of an interpolation of the fields, which are known only on a
grid. For the sake of rapidness we use a linear interpolation. This
means, that at the grid boundaries the derivatives of our field
quantities are not defined. For this reason we use a Runge--Kutta
procedure of second order accuracy in combination with an adaptive
stepsize control.

\subsection{Particle simulations: Energies and Spectra} \label{sub:spectra}

\begin{figure}[htbp]
  \centering
  \includegraphics[angle=0,width=\hsize,totalheight=\figheight,keepaspectratio]{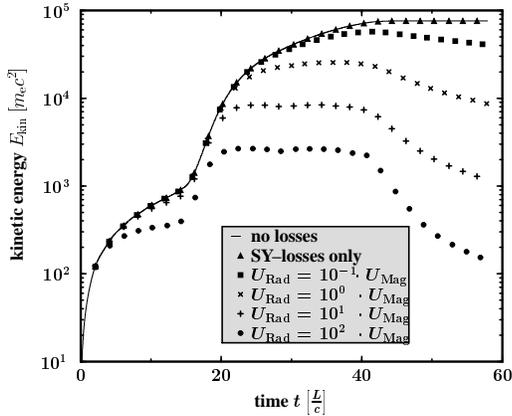}
  \caption[]{Temporal evolution of the kinetic energy of a specific
  electron with respect to different loss processes and strengths}
  \label{fig:time}
\end{figure}

In our calculations we have obtained the following results. The
SY--losses are completely unimportant in our AGN context. In
Fig.~\ref{fig:time}
one can see clearly, that the curves without any losses and with
SY--losses only are nearly identical. In the following all energies are
given in units of the rest energy of the electrons~$m_{\rm e} c^2$ and
$U_{\rm Mag}$ denotes the energy density of the magnetic field. For
IC--losses the situation is completely different. The final energies
decrease steadily with increasing photon energy density. However, even
at the extreme value of $U_{\rm Rad} \simeq 100 \cdot U_{\rm Mag}$ the
electrons gain some several tens of MeV.  Anyway, one can clearly see
that for reasonable photon energy densities ($U_{\rm Mag}\simeq
U_{\rm Rad}$) a significant part of the injected electrons are
accelerated towards GeV-energies. This can solve the injection problem.

\begin{figure}[htbp]
  \centering
  \includegraphics[angle=0,width=\hsize,totalheight=\figheight,keepaspectratio]{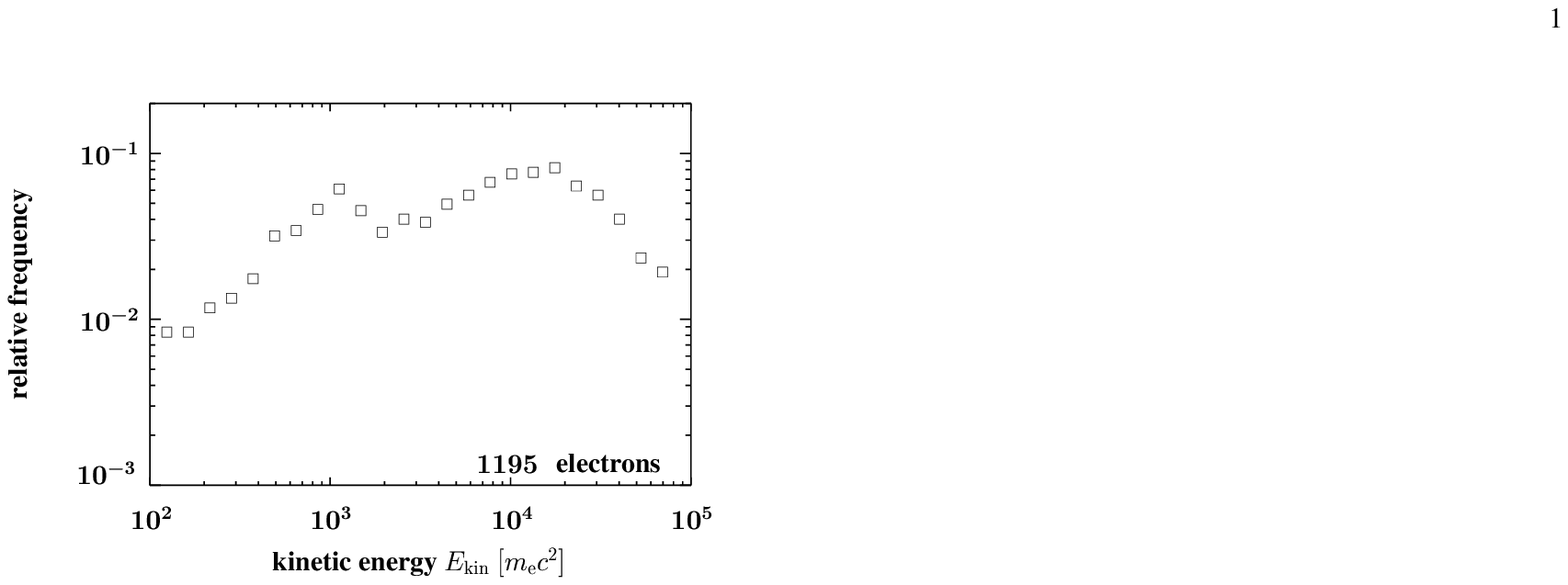}
  \includegraphics[angle=0,width=\hsize,totalheight=\figheight,keepaspectratio]{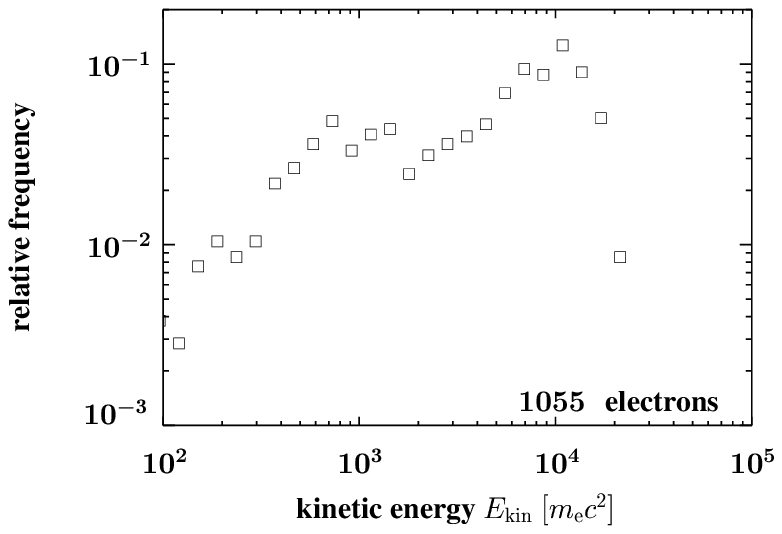}
  \includegraphics[angle=0,width=\hsize,totalheight=\figheight,keepaspectratio]{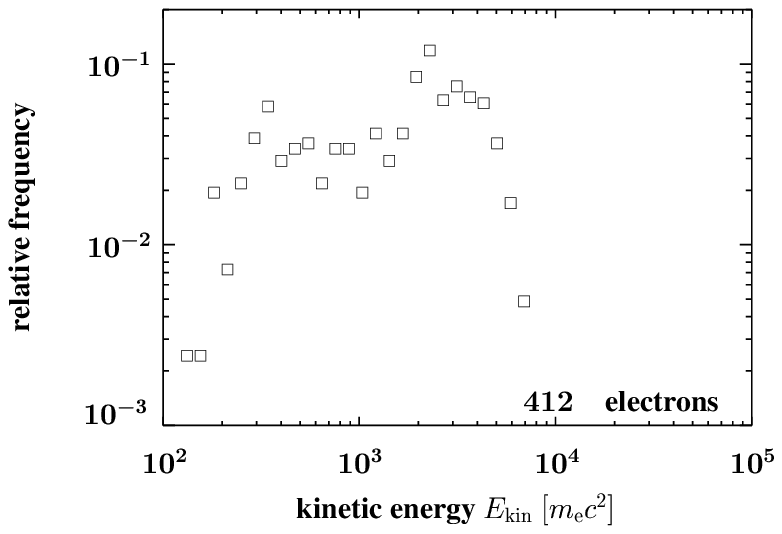}
  \caption[]{Final energy distributions of the accelerated electrons
  for three different IC--loss strengths (top:~no~losses, middle:~$U_{\rm
  Rad} = 10 \cdot U_{\rm Mag}$, bottom:~$U_{\rm Rad} = 100 \cdot U_{\rm
  Mag}$) }
  \label{fig:spectra}
\end{figure}

The final energy spectra of the accelerated electrons for different
strengths of IC--losses are shown in Fig.~\ref{fig:spectra}.
For all spectra, the electrons were injected homogeneously distributed
in the acceleration region with a velocity--distribution that
corresponds to a Maxwell-distribution of temperature~$6 \cdot 10^8
\Dim{K}$. In any case, the resulting final spectrum is a very broad
one, which goes over some decades. The effects of IC--losses are
clearly visible. The corresponding maximum of the distributions is
shifted towards lower energies with increasing photon densities. It
shifts from about~$8 \Dim{GeV}$ without energy losses to
approximately~$1 \Dim{GeV}$ at an energy density of $100 U_{\rm
Mag}$. The same tendency holds for the maximum energy achievable by
the electrons. The maximum energy gain without losses coincides with
the energy gain one would expect from the potential shown in
Fig.~\ref{fig:Potential}.  Due to the fact, that the minimum energy
gain in these distributions is quite constant with respect to a
variation of $U_{\rm Rad}$ (this results from the
$\gamma^2$--dependence of $\VV{F}_{\rm Rad}$), the distribution
becomes steeper and narrower with stronger energy losses.

\begin{figure}
  \centering
  \includegraphics[angle=0,width=\hsize,totalheight=\figheight,keepaspectratio]{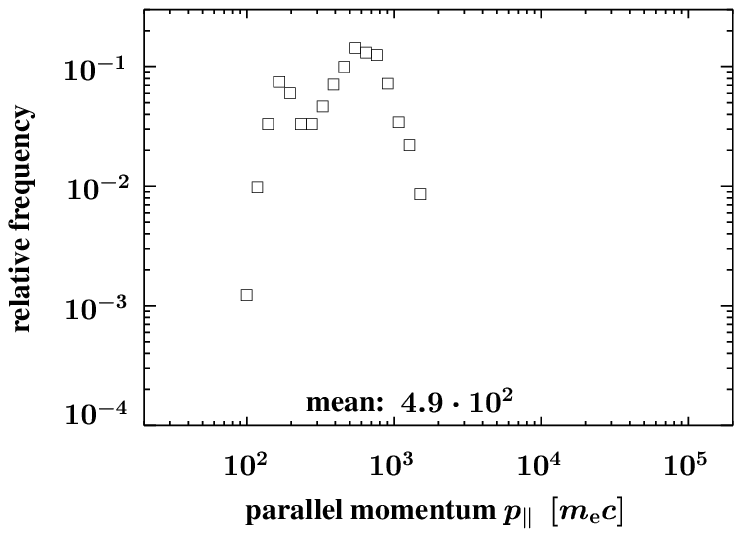}
  \includegraphics[angle=0,width=\hsize,totalheight=\figheight,keepaspectratio]{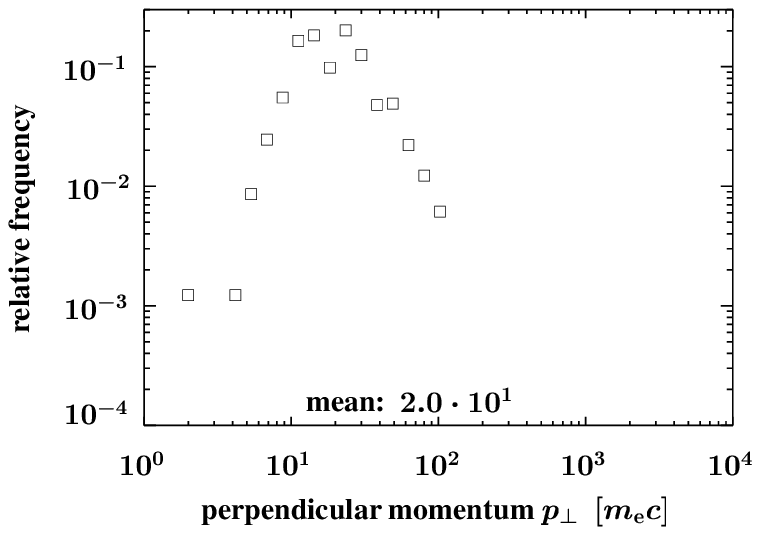}
  \caption[]{Comparison between the distribution of the final
  momentum parallel~(top) and perpendicular~(bottom) to the magnetic
  fieldlines (Note that the abscissae are different in the two figures)}
  \label{fig:momentum}
\end{figure}

Fig.~\ref{fig:momentum}
shows the distributions of the final momenta of the particles
perpendicular and parallel to the magnetic field. It is evident, that
the motion of the electrons is strongly anisotropic. They are mainly
following the magnetic field lines, whereas any motion vertical to the
field lines is suppressed. Thus, one may conclude that the accelerated
electrons represent a collection of quasi--monoenergetic beams. It is
not shown, that the movement becomes the more isotropic, the higher
the losses are. But this is straightforward, because in the loss
force used ~$\VV{F}_{\rm Rad}$ we have assumed an isotropic photon bath
which naturally tends to isotropize the electron orbits.

\begin{figure}
  \centering
  \includegraphics[angle=90,width=\hsize,totalheight=\figheight,keepaspectratio]{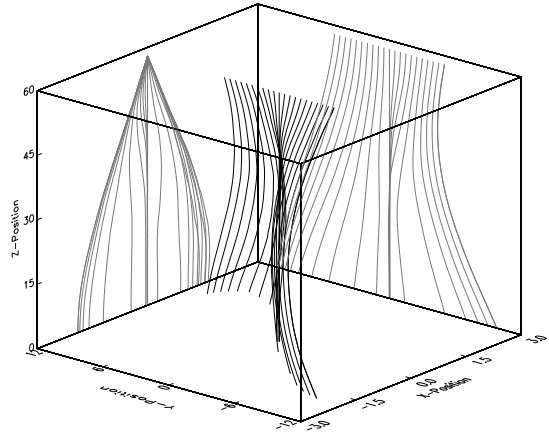}
  \caption[]{Trajectories of a chosen set of electrons. The
  projections on the walls behind, were added artificially for a
  better 3-D impression. }
  \label{fig:trajectories}
\end{figure}

In Fig.~\ref{fig:spectra} and Fig.~\ref{fig:momentum}~(top) one also
can see another characteristic feature of our field configuration. It
is quite obvious that in all four distributions there is something
like an absorption line between the two local maxima. This effect can
be explained in a natural way, if one looks directly at some chosen
trajectories of the particles. Fig.~\ref{fig:trajectories}
shows the trajectories of the electrons in combination with a
projection of the orbits on the walls behind. These projections show,
that there are two populations of electrons. One population is
accelerated towards and into the central region of reconnection,
whereas the other population is repelled. A comparison of the final
energies of the electrons with the membership in one of the two
classes shows that exactly these electrons which are forced into the
reconnection zone are accelerated the most. The members of the other
group of electrons only gain some moderate amount of energy. This
indicates that the instability of the orbits is the cause for the
appearance of the "absorption line". This effect is quite interesting,
because it represents a selection process, which leads to an even
stronger beaming of the attracted electrons and to an even stronger
anisotropy.

\section{Summary and  Discussion} \label{sec:Disc}

We addressed the question of charged particle acceleration during
magnetic reconnection processes by means of relativistic test particle
simulations.  Whereas this point is crucial for a great variety of
cosmic plasmas in this contribution we dwelled on the pre-acceleration
problem in the AGN context.  Possible further applications include
among others as different plasma systems as the terrestrial discrete
auroral arcs, the solar coronal flares, radio activity in T-Tauri
magnetospheres, non-thermal emission at the
edges of high-velocity clouds that hit the galactic plane and the
generation of electron beams in the magnetospheres of neutron stars.
The starting point for our present investigations are results obtained
by an MHD simulation study (LB). In contrast to previous work carried
out by different other groups, we were able to study particle
acceleration in large-scale non-linearly developed reconnection
electromagnetic fields rather than being restricted to the
prescription of somewhat idealized analytical field solutions.
Moreover, since for the considered parameter regime we have to expect
an intense radiation field the test particle simulations were
performed including the relevant radiative losses. A very important
question is whether charged particles can really be accelerated in the
reconnection region up to the pretty high energy values one might
deduce from the fluid treatment.  Our findings indicate that, in fact,
as expected from the fluid simulations (LB), leptons can be
accelerated in reconnection zones located in AGN coronae up to high
Lorentz factors; i.e. a significant portion of test particles gain
about the maximum energy provided by the generalized field-aligned
electric potential structures formed during the magnetic reconnection
processes we have modeled within the framework of MHD. Thus, particle
acceleration in reconnection zones may be considered as a way
out of the injection problem we face in the AGN context.

Since sheared magnetic fields can be expected as a very common
phenomena in cosmic environments like accretion disks, stellar coronae
and interstellar medium, we think that our relativistic particle
studies can be regarded as realistic with respect to the used
configuration and the dominant forces. In this contribution the
magnetic reconnection process is driven by some resistive mechanism
originating from plasma instabilities. The excited electromagnetic
oscillations serve as resistance. We note
that in plasmas which are collisionless (both no Coulomb collisions
and no turbulent wave excitation), the particle inertia presents the ultimate
source of resistivity and for the magnetic dissipation. The sheared
magnetic fields in collisionless systems evolve into very thin
filaments, in which the lifetime of the particle determines the
electrical conductivity, thereby allowing for efficient dissipation
via effective particle acceleration (Lesch and Birk 1998).

Our simulations are test--particle simulations, thus, we plan for
future studies to include, additionally, ponderomotive forces and the
back reaction of the current carried by the high energy
particles. Whether or not the latter aspect becomes important depends
on the density of the run-away electrons limited by the Dreicer
electric field (e.g. Benz 1993).

\begin{acknowledgements}
  This work was  supported by the the
  Deutsche Forschungsgemeinschaft through the grant LE 1039/3-1.

\end{acknowledgements}

{}

\end{document}